\documentclass[%
 reprint,
 amsmath,amssymb,
 aps,
]{revtex4-1}

\usepackage{graphicx}
\usepackage{dcolumn}
\usepackage{bm}
\usepackage{pictex, dcpic}

\def\SO{\hbox{SO}}
\def\SL{\hbox{SL}}
\def\SU{\hbox{SU}}
\def\Spin{\hbox{Spin}}
\def\Aut{\hbox{Aut}}
\def\id{\hbox{id}}
\def\ad{\hbox{ad}}
\def\C{\mathbb{C}}
\def\R{\mathbb{R}}
\def\one{\mathbb{I}}

\def\al{\alpha}
\def\be{\beta}
\def\ga{\gamma}
\def\ep{\epsilon}
\def\de{\delta}
\def\la{\lambda}
\def\si{\sigma}
\def\io{\iota}
\def\om{\omega}
\def\vp{\varphi}

\def\arr{\rightarrow}
\def\then{\>\Rightarrow\>\>}
\def\del{\partial}
\def\({\left(}
\def\){\right)}

\def\goth{\frak{h}}
\def\gotg{\frak{g}}
\def\slC{\frak{sl}}
\def\su{\frak{su}}

\def\[{\begin{equation}}
\def\]{\end{equation}}

\begin{document}

\preprint{APS/123-QED}

\title{Inducing Barbero-Immirzi Connections along $\SU(2)$-reductions\\
	of Bundles on Spacetime}

\author{L. Fatibene}
\email{lorenzo.fatibene@unito.it}
\affiliation{Department of Mathematics, University of Torino (Italy)}%
\affiliation{INFN- Sezione Torino; Iniz.~Spec.~Na12 (Italy)}%

\author{M. Ferraris}
 \email{marco.ferraris@unito.it}
\affiliation{Department of Mathematics, University of Torino (Italy)}%

\author{M. Francaviglia}
\email{mauro.francaviglia@unito.it}
\affiliation{Department of Mathematics, University of Torino (Italy)}%
\affiliation{INFN- Sezione Torino; Iniz.~Spec.~Na12 (Italy)}%


\date{\today}

\begin{abstract}
We shall present here a general apt technique to induce connections along bundle reductions which is different from the standard {\it restriction}. 
This clarifies and generalizes the standard procedure to define Barbero-Immirzi (BI) connection,
though on spacetime. The standard {\it spacial} BI connection used in LQG is then obtained by its spacetime 
version by standard restriction.

The general prescription to define such a {\it reduced connection} is interesting from a mathematical viewpoint and it allows a general and direct control on transformation laws of the induced object.
Moreover, unlike what happens by using standard restriction, we shall show that once a bundle reduction is given, 
then {\it any} connection induces a {\it reduced connection} with no constraint on the original holonomy as it happens when connections are simply {\it restricted}.
\end{abstract}

\pacs{Valid PACS appear here}
\maketitle


\section{Introduction}

Barbero-Immirzi (BI) connection is used in LQG to describe gravitational field on space; see \cite{Barbero}, \cite{Immirzi}. 
In standard literature it is obtained by a canonical transformation on the phase space of the spatial Hamiltonian system describing classical GR; see \cite{RovelliBook}.

Samuel argued that there is no spacetime connection which {\it restricts} to BI connection due to holonomy considerations; see \cite{Samuel}.
Thiemann claimed (see \cite{Thie2006}) that all it is needed for the theory to make sense is the definition of the connection on space, while Samuel and others would privilege spacetime objects.
Despite we partially agree with Thiemann's point of view, we have to remark that even when bundle topologies are assumed to be trivial and therefore there is no issue about objects' globality,
still transformation laws are essential for the interpretation of the theory. In these trivial situations transformation laws are not used to obtain globality, but they are used for {\it covariance}.
For the object defined to be called BI-connection it must trasform as a $\SU(2)$-connection, though transformation laws are inherited by the original spin connections and cannot be {\it imposed} at will.
Moreover, one has to define the $\SU(2)$-gauge transformations as a  subgroup of the original $\Spin(3,1)$-gauge transformations and such a subgroup must be defined canonically, i.e.~in a gauge and observer-independent fashion.

This is particularly evident when one considers that the BI connections are then described by means of their holonomy; of course holonomies are motivated and meaningful only for connections and one could not 
be satisfied with a generic spatial field which resembles a $\SU(2)$-connection but has different transformation laws.
If the action of the gauge group is modified then the holonomies are not necessarily  gauge covariant quantities any longer. 
On the other hand, if gauge covariance is abandoned the hole argument (see \cite{RovelliBook}) is compromised and the physical observables of the theory (together with its interpretation) are compromized, too. 

For these reasons we have investigated a possible construction to define BI connection keeping gauge covariance under full control; see \cite{R1}.
The construction is based on the existence of a $\SU(2)$-reduction of the original principal spin bundle $P$. A {\it $\SU(2)$-reduction} is 
a pair $({}^+P, \io)$ where ${}^+P$ is a $\SU(2)$-bundle
and $\io:{}^+P\arr P$ a (vertical) principal morphism with respect to the canonical group embedding $i:\SU(2)\arr \Spin(3,1)$:
\[
\begindc{\commdiag}[1]
\obj(110,80)[+P2]{$^+P$}
\obj(180,80)[hP3]{$P$}
\obj(110,30)[M2]{$M$}
\obj(180,30)[M3]{$M$}
\mor{+P2}{M2}{}
\mor{hP3}{M3}{}
\mor{+P2}{hP3}{$\io$}
\mor{M2}{M3}{}[\atleft, \solidline] \mor(110,33)(180,33){}[\atleft, \solidline]
\enddc
\]
In standard situations,  when spacetimes are required to allow global Lorentzian metrics and global spinors (that is equivalent to require that first and second Stiefel-Whitney classes vanish)
such a reduction can be shown to exist always (see \cite{Antonsen}) with no further topological obstruction.
For a simplified situation, when we can imagine the spin bundle $P$ to be trivial, the reduction always exists and the reduced bundle ${}^+P$ is also trivial.

The $\SU(2)$-reduction defines a {\it canonical} embedding of $\SU(2)$-gauge transformations (namely, $\Aut({}^+P)$) into the $\Spin(3,1)$-gauge  transformations (namely, $\Aut(P)$).
One can now consider a spin connection $\om$ on $P$. 
This cannot always be {\it restricted} to ${}^+P$. To be able to restrict the connection $\om$ to the sub-bundle $\io({}^+P)\subset P$, $\om$-horizontal spaces must happen to be tangent to the sub-bundle itself.
Of course, this is a condition on $\om$ for it being restrictable; a trivial necessary condition for this is that the holonomy of the original connection $\om$ happens to get value in the subgroup $\SU(2)\subset \Spin(3,1)$
in the first place. Hence there are spin connections that cannot be restricted (see \cite{Samuel}; we thank Smirnov for addressing our attention on this point \cite{Smirnov}).

In \cite{R1} we proposed a {\it different} prescription to induce a $\SU(2)$-connection $A$ on ${}^+P$ out of the spin connection $\om$ on $P$; despite this presciption is not canonical 
(and below we shall describe exactly in which sense it is not) it is {\it generic}; all spin connections $\om$ induce a {\it reduced} connection $A$ on ${}^+P$, in particular with no restriction on holonomies.

The construction is possible if an algebraic relation (a {\it reductive splitting}, see Section 2 below) between the involved groups holds. For the groups of interest for LQG (i.e.~$i:\SU(2)\arr \SL(2,\C)$) this reductive splitting always exists and thence the construction is possible. This supports the results in \cite{R1}.
However, one can show that in other dimensions or other groups the splitting is not reductive. 
Thence the construction for BI connection is not general and cannot be extended to generic situations.
This kinematical issue adds up to the fact that Holst's action principle is characteristic of dimension $4$ and in a generic situation some other dynamics should be introduced.

To summarize, we showed in \cite{R1} that one can define a $\SU(2)$-connection on ${}^+P$, i.e.~over spacetime. 
This connection can be then restricted to space to obtain the standard BI connection. 
However, the spacetime reduced $\SU(2)$-connection is not the restriction of a spin connection on spacetime and its holonomy is not necessarily dictated by the original spin connection (which therefore is not required to be in $\SU(2)$ as argued instead in \cite{Samuel} and \cite{Smirnov}).

The construction shows that the holonomy of BI connection encodes the holonomy of the original spin connection in a non-trivial way. Further investigation is needed to understand this encoding in detail. 
Such issues have to be clarified for example to investigate the semiclassical limit which in LQG is far from clear.

This paper is organized as follows: 
in Section $2$  we shall define the reduction prescription from a more general point of view with respect to what we did in \cite{R1}.
In Section $3$ we shall obtain the BI prescription defined in \cite{R1} from our new and more general point of view.

\section{Induced Connections along Reductions and Reductive Algebras}

In this Section we shall consider the algebraic structures that enable us to reduce the connections.
Let us consider a principal bundle $P$ with group $G$ and a subgroup $i: H\arr G$.
Let us then assume and fix any {\it $H$-reduction} $(Q, \io)$ given by 
\[
\begindc{\commdiag}[1]
\obj(110,80)[hQ3]{$Q$}
\obj(180,80)[hP3]{$P$}
\obj(110,30)[M2]{$M$}
\obj(180,30)[M3]{$M$}
\mor{hQ3}{M2}{}
\mor{hP3}{M3}{}
\mor{hQ3}{hP3}{$\io$}
\mor{M2}{M3}{}[\atleft, \solidline] \mor(110,33)(180,33){}[\atleft, \solidline]
\enddc
\]

The existence of such a reduction usually imposes topological conditions on the spacetime. 
As we already remarked in Section 1 in the standard situation of $G=\Spin(3,1)$ and $H=\SU(2)$ the bundle reduction is automatically ensured 
by standard physical requirements (essentially by existence of global spinors).

In a more general case one should discuss the condition for this reduction to exist, usually rephrasing it in terms of vanishing of cohomology classes.
In the standard case it amounts to the vanishing of the third Stiefel-Whitney class; see \cite{Antonsen}.
In this paper we wish to show that in order to define a $H$-connection one also needs the relevant groups to obey an algebraic condition, namely that $H$
is {\it reductive} in $G$ as defined hereafter.

This aspect can be easily discussed for general groups (when the existence of bundle reduction is assumed and it corresponds to a well-understood structure). Then we shall show that in the standard case ($G=\Spin(3,1)$ and $H=\SU(2)$) the groups are automatically reductive.

The group embedding  $i:H\arr G$ induces an algebra embedding $T_e i: \goth\arr \gotg$.
Let us define the vector space $V=\gotg/\goth$ so to have the short sequence of vector spaces
\[
\begindc{\commdiag}[1]
\obj(10,20)[oL]{$0$}
\obj(50, 20)[h]{$\goth$}
\obj(100, 20)[g]{$\gotg$}
\obj(150, 20)[V]{$V$}
\obj(190, 20)[oR]{$0$}
\mor{oL}{h}{}
\mor{h}{g}{$T_e i$}
\mor{g}{V}{$p$}
\mor{V}{oR}{}
\cmor((150,15)(145,8)(125,6)(105,8)(100,15))
	\pup(125,1){$\Phi$}[\atleft, \dasharrow]
\enddc
\]
where $\Phi: V\arr \gotg$ is a sequence splitting (i.e.~$p \circ \Phi= \id_V$) which always exists for sequences of vector spaces. 
Accordingly, one has  $\gotg\simeq \goth\oplus \Phi(V)$.

We say that $H$ is {\it reductive} in $G$ if there is an action $\la: H\times V \arr V$ such that $\ad(h)(\Phi(v))\equiv \Phi\circ \la(h, v)$
where $\ad: H\times \gotg \arr \gotg$ is the restriction to the subgroup $H$ of the {\it adjoint action} of $G$ on its algebra $\gotg$; see \cite{KobaNu}, \cite{Gatto}, \cite{GM}. 
In other words, the subspace $\Phi(V)\subset \gotg$ is invariant with respect to the adjoint action of $H\subset G$ on the algebra $\gotg$.

Let us stress that the vector subspace $\Phi(V)\subset \gotg$ is not required to be (and often it is not) a subalgebra; accordingly, one is not choosing any group splitting
$G= H\times K$ (as for example it happens (incidentally) in the case of the (anti)selfdual decomposition $\Spin(4)= \SU(2)\times \SU(2)$). 
A group splitting (and the corresponding projection) is not at all used; one just needs the group embedding $i:H\arr G$.

We shall show hereafter that a  bundle $H$-reduction $\io:Q\arr P$ with respect to a subgroup $H$ reductive in $G$ is enough to allow that 
{\it each} $G$-connection $\om$ on $P$ induces an $H$-connection on $Q$, which will be called the {\it reduced connection}.

Let us consider a $G$-connection $\om$ on $P$ locally given by
\[
\om= dx^\mu\otimes \(\del_\mu -\om^A_\mu(x)\rho_A\)
\]
where $\rho_A$ is  the pointwise basis for vertical right invariant vector fields on $P$ associated to a basis $T_A$ of the Lie algebra $\gotg$; see \cite{Book} for notation.

Resorting to the algebra splitting one can consider an {\it adapted basis} $T_A=(T_i, T_\al)$, $T_i$ being a basis of $\goth$ and $T_\al$ a basis of $\Phi(V)$.
The corresponding basis of vertical right invariant vector fields on $P$ splits as $\rho_A=(\rho_i, \rho_\al)$.

In view of  the reductive splitting of the algebras, for any $H$-gauge transformation $\vp:U\arr H$,  one has
\[
(\rho'_i, \rho'_\al)\equiv\rho'_A= \ad^B_A(\vp)\rho_B \equiv (\ad_i^j(\vp)\rho_j, \la_\al^\be(\vp) \rho_\be)
\] 
Accordingly, the $G$-connection can be splitted as
\[
\om= dx^\mu\otimes \(\del_\mu -\om^i_\mu(x)\rho_i\) \oplus (-\om^\al_\mu(x) dx^\mu\otimes\rho_\al)
\]
Since $\rho_i$ transform with respect to the adjoint representation of $H$ and $\rho_\al$ transform wrt to the representation $\la$, then the quantities
\[
A= dx^\mu\otimes \(\del_\mu -\om^i_\mu(x)\rho_i\)
\quad
K=-\om^\al_\mu(x) dx^\mu\otimes\rho_\al
\]
are (modulo trivial and canonical isomorphisms) an $H$-connection on $Q$ and a vector valued $1$-form on $Q$, respectively.

In the following Section we shall show how the standard BI connection can be obtained in this framework as done in \cite{R1}.

Let us stress that, once the $H$-reduction is assumed and the corresponding splitting is shown to be reductive, then {\it all} connections $\om$ of $P$ 
induce a $H$-connection $A$ on $Q$, in particular with no holonomy constraints. 

As argued in \cite{Smirnov}, torsionless connections obey severe constraints on possible holonomies they can have; see \cite{HolClass}, \cite{HolClassExc}.
These results do not directly apply to gauge connections (and spin connections in particular); however, when a frame is considered, as it is done in LQG, spin connections induce also spacetime connections 
which {\it are} in fact constrained in their possible holonomies, so that one could eventually consider this as a constraint on the holonomy of the original spin connection.
Since GR field equations imply  torsionless spin connections, then a potential issue can be considered:

\begin{itemize}
\item[]can torsionless $\Spin(1,3)$-connections (among which all solutions of GR) induce $\Spin(3)$-connections when the holonomy group $\Spin(3)$ is forbidden by the classification?
\end{itemize}

The answer is in the negative if $\Spin(3)$-connections are induced by {\it restriction}. 
But it is in the positive if $\Spin(3)$-connections are induced by {\it reduction} as above.

Of course, one could argue that the existence of bundle reduction and  the reductive splitting is  a constraint equivalent to the one on the holonomies.
However, this is not the case; one can consider the subgroup $i:\SU(2)\arr \Spin(3,1)$ which is in fact {\it reductive} (as we shall show below).
If the spin bundle $P$ considered is  trivial then there is no topological obstruction to the existence of the reduction $\io:{}^+P\arr P$.
In this situation all hypotheses about the prescription for reduced connections are satisfied 
and {\it each} $\Spin(3,1)$-connection induces a reduced $\SU(2)$-connection, included the torsionless connections which cannot be {\it restricted} in view of the constraints on holonomy.

\section{An Example: $i:\SU(2)\arr \Spin(3,1)$}

The group $\Spin(3,1)$ is isomorphic to $\SL(2,\C)$ which is a sort of {\it complexification} of $\SU(2)$ that is identified 
accordingly as a real section $i:\SU(2)\arr \SL(2, \C)$.

The corresponding algebra of $\slC(2,\C)$ is spanned (on $\R$) by $(\tau_i, \si_i)$ where $\si_i$ are standard Pauli matrices and $\tau_i=i \si_i$.
An element of $\slC(2,\C)$ is thence in the form $\xi= \xi^i_{(1)} \tau_i+ \xi^i_{(2)} \si_i$ and the algebra embedding is given by
\[
T_ei: \su(2) \arr \slC(2,\C): \xi^i \tau_i \mapsto  \xi^i \tau_i
\]
The quotient $V$ is spanned by $\si_i$ and 
the splitting of the algebra sequence can be fixed as
\[
\Phi: V\arr \slC(2,C): \si_i \mapsto \si_i + \ga \tau_i
\qquad( \ga\in \R)
\]
 which is in fact always transverse to $\su(2)\subset \slC(2,\C)$.

One can easily show that such a splitting is reductive and the representation $\la: \SU(2)\arr \SO(3)$ coincides with the standard covering 
map exhibiting the group $\SU(2)$ as the double covering of the orthogonal group $\SO(3)$ on space.

{
In fact one can consider $S= a_0\one + a^i\tau_i\in \SU(2)$, which is obtained by $a_0, a^i\in \R$ with $(a_0)^2 + |\vec a|^2=1$ and set $\ga \tau_k + \si_k= e_k$. Then one can show that 
\[
\begin{array}{rl}
S e_k S^{-1} = &  
\(\((a_0)^2    -  |\vec a|^2\) \de^j_k -2a_0a^i\ep_{ik}{}^j +2  a^\cdot_k a^j  \) e_j   =\\
=& \la^l_k(S) e_l\\
\end{array}
\label{CoveringMap}
\]
This shows how $Se_k S^{-1}\in V$, hence the splitting is reductive and the representation $\la$ is given by
\[
\la: \SU(2)\times V \arr V: (S, e_k)\mapsto \la^l_k(S) e_l
\]
where in view of (\ref{CoveringMap}) one has
\[
\la^j_k(S)=\((a_0)^2    -  |\vec a|^2\) \de^j_k -2a_0a^i\ep_{ik}{}^j +2  a^\cdot_k a^j 
\]
}

Let us also stress that $\Phi(V)$ in this case is not a subalgebra.
{\small
It is sufficient to show that $V$ is not closed with respect to commutators.
For example (assuming $\ga\not=0$) one has:
\[
\begin{array}{rl}
[\si_1+&\ga\tau_1, \si_2+\ga\tau_2]= [\si_1, \si_2] + 2\ga [\tau_1, \si_2] + \ga^2[\tau_1, \tau_2]=\\
=&2(\ga^2+1) \tau_3 -4\ga \si_3 = 2\ga\( \frac{1}{\ga}\tau_3 - \si_3\) -2\ga \(\si_3+\ga \tau_3\)\\
\end{array}
\]
The result is not in $\Phi(V)$ unless one has $-\frac{1}{\ga}=\ga$ (i.e.~ $\ga^2+1=0$).
}

The basis $\si_{ab}$ of vertical right invariant vector fields on $P$ is given by the following identification with the algebra
\[
\begin{array}{ccc}
-4\si_{12}= \tau_3  &   4\si_{13}= \tau_2  &   -4\si_{23}= \tau_1 \\
4\si_{01}= \si_1  &   4\si_{02}= \si_2  &   4\si_{03}= \si_3 \\
\end{array}
\label{AlgebraGeneratorsSU2}
\]
as one can check by computing commutators of fields $\si_{ab}$ (see Appendix $A$ for notation). 
Hence the basis of $\Phi(V)$ is $e_k=  4\(\si_{0k} +\frac{\ga}{2} \ep_k{}^{ij}\si_{ik} \)$, i.e.
\[
\begin{array}{c}
e_1= 4\(\si_{01} + \ga \si_{23}\) \\
e_2= 4\(\si_{02}-\ga \si_{13}\) \\
e_3=  4\(\si_{03}+\ga \si_{12}\)\\
\end{array}
\]

Then we can split a generic connection
\[
\begin{array}{rl}
\om=& dx^\mu\otimes\( \del_\mu - \om^{ab}_\mu\si_{ab}\)=  \\
=&dx^\mu\otimes\( \del_\mu - \om^{ij}_\mu\si_{ij}- 2\om^{0i}_\mu\si_{0i}\)=\\
=&dx^\mu\otimes\( \del_\mu - \om^{ij}_\mu\si_{ij}- 2\om^{0i}_\mu(\si_{0i} \pm\frac{\ga}{2}\ep_i{}^{jk} \si_{jk})\)=\\
=&dx^\mu\otimes\( \del_\mu - (\om^{jk}_\mu+\ga\om^{0i}_\mu\ep_i{}^{jk}) \si_{jk}\) - \frac{1}{2}\om^{0i}_\mu e_i\\
\end{array}
\]
Hence one can define
\[
A^i_\mu=\frac{1}{2}\ep^i{}_{jk}A^{jk}_\mu =\frac{1}{2}\ep^i{}_{jk}\om^{jk}_\mu+ \ga\om^{0i}_\mu 
\qquad
K^i_\mu= - \frac{1}{2}\om^{0i}_\mu
\]
According to the general theory, $A^i_\mu$ is a $\SU(2)$-connection and $K^i_\mu$ is a Lie algebra valued $1$-form;
this can be easily seen directly as was done by quite complicated computation in \cite{R1}.
Reductive splittings provide a clear and simple way to keep transformation laws (and globality) under full control.
It also amounts to an algebraic fact in group theory which can be easily considered in a generic situation.

Now that we have reproduced the results of \cite{R1}, we are ready to show that the ones considered are the {\it only} reductive splittings.
A generic splitting is in fact $\Phi:V\arr \slC(2, \C): \si_i \mapsto \si_i+ \be_i^j \tau_j$. 
If one imposes reductivity one easily finds the condition
\[
\be_i^m\de_{jk}= \de_{lj} \de_i^m \be^l_k
\label{beCond}
\]
that is satisfied if and only if $\be_i^j=\ga \de_i^j$.

{
In fact let us set $e_k= \si_k + \be_k^i\tau_i$.
Following the line of the proof of reductivity given above one can easily show that
\[
\begin{array}{rl}
S \cdot e_k \cdot S^{-1}=& \la_k^j (S) e_j + 2a^i a^j (\de_i^m \be_k^l\de_{lj} - \be_j^m\de_{ki}) \tau_m  +\\
&+ 2a_0a^j (\be_k^l\ep_{lj}{}^m -\ep_{kj}{}^n \be_n^m) \tau_m\\
\end{array}
\] 
Since the span of $(\tau_n, n=1,2,3)$ is transverse to $\Phi(V)$ which is spanned by $(e_k: k=1,2,3)$ the extra terms must vanish for all $S\in \SU(2)$.

Hence one must have
\[
\begin{cases}
\de_{(i}^m \de_{j)l} \be_k^l = \be_{(j}^m\de_{i)k} \\
\be_k^l\ep_{lj}{}^m =\ep_{kj}{}^n \be_n^m
	\quad\then
	\ep^i{}_h{}^j \(\be^m_{[j} \de_{i]k} - \de^m_{[j} \de_{i]l} \be^l_k\)=0\\
\end{cases}
\]
That implies
\[
\be^m_{j} \de_{ik} = \de^m_{j} \de_{il} \be^l_k
\]
which proves equation (\ref{beCond}).
By tracing (\ref{beCond}) wrt the indices $(im)$ one has
\[
\be \de_{jk}= 3\de_{lj} \be^l_k
\quad\then
\be^j_k= \frac{\be}{3}  \de^j_k
\]
}

Let us stress that any other reductive splitting had we found, it would have allowed other connections like BI connections, though enumerated by a {\it matrix Immirzi parameter} $\be^i_k$, that in principle should have allowed alternative LQG-like formalisms. Thus we believe the negative result (which to the best of our knowledge is new in literature)
is important when discussing (non-)uniqueness of LQG approach.

\section{Conclusions and Perspectives}

We showed that BI-connections can be properly understood in terms of bundle reductions along reductive group splittings.
Let us stress that the understanding of the geometric origin of BI connection is necessary as far as it guarantees a control on global properties (passive viewpoint) and equivalently on gauge-covariance with respect to general $\SU(2)$-gauge transformations (active viewpoint).
These aspects are equivalent and they are necessary, e.g., to guarantee gauge covariance of holonomies which is a fundamental motivations for choosing holonomies (or spin networks) to parametrize the physical degrees of freedom in LQG.

Although one cannot show this mechanism to be strictly necessary, we stress that it is currently the only known mechanism which allows to control these properties.
To claim that an object is a connection one necessarily needs to define a suitable principal bundle and to show that components transform as expected under the general automorphisms of the principal bundle which play the role of gauge transformations. Without defining the principal bundle one cannot even define what global general $SU(2)$-gauge transformations are!

To summarize, the standard BI connection is not the spatial restriction of a spacetime spin connection.
It is instead the restriction {\it of the reduction} of a spacetime spin connection and the restricted spacetime connection is 
the spacetime counterpart of the spatial BI connection, though it is not a $\Spin(n,1)$-connection.

While restricting, the restricted holonomy has to agree with the holonomy of the original connection, when the new connection is defined 
by reduction the new holonomy undergoes a projection procedure which has a potential impact on the holonomy. While the projection is explicit and clear
at the level of the equation defining parallel transport, we are not yet able to trace the effect on the holonomy groups.
Although we may not be able to predict and control the change of holonomy produced by restrictions which was exhibited by Samuel, his all argument rests on the assumption that the BI connection should maintain the same holonomy of the original spin connection. Which is not the case when restricting. 

Also the argument based on Berger classification is based on the assumption the BI connection must be defined by restriction. Once one realizes that it is instead defined by reduction, the argument does not stand. 

In view of these results, the BI connection $A$ (together with the field $K$) is an equivalent description of the spin connection $\om$, in the technical sense that the map $\om\mapsto (A, K)$ is a bijection. The fact that the holonomy of $\om$ is non-trivially encoded by the BI connection $A$ has to be accepted and
it can be conjectured as a motivation for the holonomy of $A$ to be easier or better suited for quantization procedure. We still cannot grasp enough details to prove such a  conjecture, still if the BI holonomies were simply a restriction of the spin holonomies one could ask why one should prefer BI over spin connections!
On the other hand, understanding in detail this coding seems to be essential to discuss semiclassical limit of LQG.

Let us finally remark that here we are  dealing uniquely with kinematics. Our analysis does not depend on the dynamics. We discussed dynamics in the standard case in \cite{R2}. In the standard case the $K$ field is determined algebraically by the triad, as it is known also from the Hamiltonian constraint analysis.
However, here we are not discussing these aspects.

Further investigation will be devoted to see whether non-trivial reductive splittings exist in  dimensions different from $m= 4$ or different signatures.
Each of such reductive splitting would allow to define a reduced connection similar to the standard BI connection, provided that one first discusses existence of  the relevant bundle reduction.
If they exist, then one will be able to  study the dynamics of higher-dimensional gravity along the lines of \cite{R2}.
This would be possible using the Holst dynamics as written in \cite{Uni1} or the modified dynamics (the ones equivalent to $f(R)$ models) as in \cite{Uni2}.

If there is no other reductive splitting $i:\Spin(n)\arr \Spin(n, 1)$ other than for the standard case $n=3$, then, independently of the existence of the relevant bundle reductions, this would seriously question the possibility of a BI-like approach in other dimensions.
To be honest, also in this negative scenario, we are not able to exclude the possibility of defining a $\Spin(n)$-connection by some other construction which does not rely on reductive splittings. Still we remark that {\it currently} there is no way to control globality of BI conditions different from the one presented here.

Let us also remark that BI connection has not been shown to be necessary to LQG. The standard current approach to LQG uses BI (and hence it is important to control its global properties) but other approaches have been investigated; see \cite{Alexandrov}, references quoted therein as well as  \cite{Livine}. 

\medskip
\section*{Appendix: Commutators of $\si_{ab}$}

A pointwise right-invariant basis for vertical vector fields on a principal $\Spin(\eta)$-bundle $P$
is induced by a frame $e:P\arr L(M)$ locally represented by the matrices $e_a^\mu$ in the form (see \cite{Book})
\[
\si_{ab}= \eta_{c[b}^{\phantom\mu} e_{a]}^\mu \frac{\del}{\del e_c^\mu}
\]
One can easily prove that the commutators are
\[
[\si_{ab}, \si_{cd}] =\( \eta_{c[a}\de_{b]}^e\de_d^f - \eta_{d[a}\de_{b]}^e\de_c^f  \) \si_{ef}
\label{CommutatorsSU2}
\]

In dimension $4$ the indices run in the range $a, b= 0,..3$ and one can set 
\[
	\hat\si_i:=4 \si_{0i}
\qquad
	\hat\tau_i=-2\ep_i{}^{jk}\si_{jk}	\quad(\then \si_{jk}=-\frac{1}{4}\ep_{jk}{}^i \hat\tau_i)
\]
The commutators (\ref{CommutatorsSU2}) specify to
\[
[\hat\si_{i}, \hat\si_{j}] =  2 \ep_{ij}{}^k \hat\tau_{k}
\quad
[\hat\si_{i}, \hat\tau_{j}] = -2\ep_{ij}{}^k\hat\si_k
\quad
[\hat\tau_{i}, \hat\tau_{j}] = -2\ep_{ij}{}^k\hat\tau_k
\]
which accounts for the identification of vertical vector fields with algebra generators given by (\ref{AlgebraGeneratorsSU2}).

\begin{acknowledgments}
We wish to thank L.~Gatto, M.~Godina and P.~Matteucci for discussions about reductive splittings
and  A.~Smirnov for having pointed out to us the argument based on holonomy constraints.

We also thank C.~Rovelli for discussion about BI-connections.

We also acknowledge the contribution of INFN (Iniziativa Specifica NA12) the local research project 
{\it Leggi di conservazione in teorie della gravitazione classiche e quantistiche} (2010) of Dipartimento di Matematica of University of Torino (Italy).
\end{acknowledgments}

\end{document}